\documentclass[11pt]{article}
\usepackage[margin=1in]{geometry}
\usepackage{graphicx}
\usepackage{subcaption}
\usepackage{caption}
\usepackage{bm}

\usepackage{amsmath,amssymb}

\usepackage{changepage}

\usepackage[utf8x]{inputenc}

\usepackage{textcomp,marvosym}

\usepackage{cite}

\usepackage{nameref,hyperref}

\usepackage[right]{lineno}

\usepackage{microtype}
\DisableLigatures[f]{encoding = *, family = * }

\usepackage[table]{xcolor}

\usepackage{array}

\providecommand{\Ro}{\mathcal{R}_0}

\begin{document}
\vspace*{0.2in}

\begin{flushleft}
{\Large
\textbf\newline{Changing Burial Practices Explain Temporal Trends in the 2014 Ebola Outbreak} 
}
\newline
\\
Michael A.L. Hayashi \textsuperscript{1*},
Marisa C. Eisenberg \textsuperscript{1,2},
\\
\bigskip
\textbf{1} Department of Epidemiology, School of Public Health, University of Michigan, Ann Arbor, MI, USA
\\
\textbf{2} Department of Mathematics, University of Michigan, Ann Arbor, MI, USA
\\
\bigskip

* mhayash@umich.edu

\end{flushleft}
\section*{Abstract}
    \paragraph*{Background:} The 2014 Ebola outbreak in West Africa was the largest on record, resulting in over 25,000 total infections and 15,000 total deaths.  Mathematical modeling can be used to investigate the mechanisms driving transmission during this outbreak -- in particular, burial practices appear to have been major source of infections.
    \paragraph*{Methodology/Principal Findings:} We developed a multi-stage model of Ebola virus transmission linked to a game-theoretic model of population burial practice selection.  We fit our model to cumulative incidence and mortality data from Guinea, Liberia, and Sierra Leone from January 2014 to March, 2016.  The inclusion of behavior change substantially improved best fit estimates and final size prediction compared to a reduced model with fixed burials.  Best fit trajectories suggest that the majority of sanitary burial adoption occurred between July, 2014 and October, 2014.  However, these simulations also indicated that continued sanitary burial practices waned following the resolution of the outbreak.
    \paragraph*{Conclusions/Significance:} Surveillance data from the 2014 outbreak appears to have a signal of changes in the dominant burial practices in all three countries.  Increased adoption of sanitary burials likely attenuated transmission, but these changes occurred too late to prevent the explosive growth of the outbreak during its early phase.  For future outbreaks, explicitly modeling behavior change and collecting data on transmission-related behaviors may improve intervention planning and outbreak response.

\section*{Author summary}
Changes in burial practices in West Africa during the 2014 Ebola outbreak may have slowed the expansion of cases by preventing transmission from corpses to healthy family members.  In particular, acceptance of WHO sanitary burial procedures may have been aided by increasing awareness of the magnitude of the outbreak.  However, it has been difficult to quantify the impact of these changes post hoc.  We developed a mathematical model of Ebola transmission that integrates game theory to capture the transition to sanitary burials over time.  Our analysis indicates that this transition may explain the outbreak patterns observed in surveillance data for Guinea, Liberia, and Sierra Leone.  In particular, the adoption of sanitary burials may have attenuated transmission, but only after the peak growth phase of the epidemic.  However, this attenuation appears to have substantially reduced the final case burden.  It may be important, therefore, to monitor behavioral adaptation during outbreaks in order to improve response and control efforts.

\section*{Introduction}
Ebola virus disease (EVD) is a serious and frequently lethal infection often identified with the profuse internal and external bleeding observed in late stage patients \cite{world2016ebola,cdc2016ebola}.  First identified in 1976, EVD has historically been restricted to small, self-limiting outbreaks \cite{world2016ebola,cdc2016ebola}.  However, in 2014 West Africa experienced the largest recorded epidemic with over 25,000 total cases and 15,000 total deaths primarily distributed among Guinea, Liberia, and Sierra Leone.  Unlike prior occurrences of EVD, the 2014 outbreak reached urban centers, amplifying its transmission potential.  In addition, regional infrastructure was ill equipped to contain the epidemic, while widespread poverty and political instability further complicated control measures \cite{alexander2014ebola, dallatomasina2015ebola}.  

The magnitude of the 2014 outbreak sparked a substantial amount of interest among mathematical epidemiologists, and a variety of models were developed to test control measures and forecast the trajectory in West Africa \cite{rivers2014modeling, yamin2015effect, fisman2014early, lewnard2014dynamics}.  However, early projections were criticized for frequently overestimated the final size of the outbreak, leading to a debate regarding the uses and effectiveness of mathematical models in outbreak situations \cite{butler2014models, rivers2014ebola}.  Much of the initial forecasting error appears to be due to the lack of adjustment for the reporting rate and population at risk in the output equations.  Including such a correction factor improves the accuracy of short and medium-term forecasting \cite{eisenberg2015ebola}.  Nevertheless, the reporting rate/population at risk adjustment does not mechanistically capture behavior changes that may be important for long-term prediction \cite{funk2014ebola,richards2015social} and intervention planning.  In particular, we concentrate on funerals as a transmission route and potential source of behavioral dynamics.

Due to the pathophysiology of EVD and its high mortality rate, regional burial practices fell under scrutiny as a major source of transmission.  Recent studies have indicated that Ebola virus can remain viable up to seven days postmortem \cite{prescott2015postmortem}, with high concentrations of pathogen remaining in body fluids.  Traditionally, burials in West Africa involve a substantial amount of direct contact between the family of the deceased and the cadaver itself.  Specifically, these practices include touching, washing, and kissing the corpse, resulting in a high probability of infection.  Indeed, a single funeral in Guinea resulted in 85 new cases of EVD \cite{victory2015ebola}.  The deeply traditional nature of these burial practices posed a challenge to public health professionals attempting to reduce funeral transmission.  Sanitary procedures such as sterilization, bagging, and disposal of corpses were viewed as an affront to the deceased and their family \cite{nielsen2015improving}, leading to ``safe and dignified" burial initiatives \cite{WHO2014burial}.  Of the three primary countries, Liberia instituted mandatory cremations, leading some burials to be conducted secretly.  In spite of these challenges, sanitary burials have been cited as a key factor in containing the outbreak \cite{team2015west,pandey2014strategies}.

We would expect the shift in burial practices from traditional to safe and dignified methods to have a measurable impact on the course of the outbreak.  To capture the effect of these changes in behavior, it is necessary to address the feedback between EVD transmission and burial practices.  The primary incentive to adopt sanitary burial techniques is likely related to the degree of EVD transmission, in this case the contribution to morbidity and mortality from traditional burials.  Thus as an EVD outbreak grows, sanitary burials may become more common, reducing the force of infection.  We explicitly model this feedback relationship using a population level game theoretic model of burial practices coupled to a compartmental model of Ebola transmission.  This approach is similar to the influenza and sexually transmitted infection models of Bauch and others \cite{bauch2005imitation,reluga2006evolving,reluga2010game,chen2004rational}.  Discrete individual simulations can also be used to address changing behavior, however such models can be computationally expensive, difficult to analyze, and may often be reframed through the lens of game theory.  The deterministic model presented here is relatively simple to implement and enables us to make inferences about behavioral processes from traditional surveillance data.  In addition, the behavior-disease framework is straightforward to extend to multiple or alternative behavioral mechanisms.

\section*{Methods}
\subsection*{Model}
\label{subsec:model}

\paragraph{Burial practices}
By definition, individuals cannot make a choice about the manner of their own burial at the time it occurs.  However, it is possible to specify what type of burial would be preferable for family members and others in the community.  In aggregate, we assume that these preferences determine the distribution of burial practices.  Individual choices exist in a feedback cycle with social or cultural norms about burial.  Individuals are influenced by the dominant burial practice, and continued adherence reinforces the norm.  We model this feedback using evolutionary game theory \cite{smith1982evolution,nowak2006evolutionary}.  Like traditional game theory, we specify the set of choices individuals can make and the abstract payoff associated with each potential choice.  However, evolutionary game theory is concerned with the dynamics of behavior in large populations of interacting individuals, which is appropriate to our analysis of changes in behavioral practices over the course of an outbreak.

As noted above, individuals choose whether they prefer a traditional or a sanitary burial.  Each type of burial incurs a cost that reflects perceived advantages of the other type of burial.  Individuals who prefer sanitary burials may face decreased social acceptance for acting contrary to the dominant cultural practice.  However, if the contribution of burials to transmission is recognized, traditional burials may begin to be perceived as dangerous.  These factors inform our payoff functions.  

The payoff $u_T$ for traditional funerals is assumed to be a decreasing function of I:
\begin{equation}
u_T(I) = -\rho I,
\end{equation} 
where $I$ is the total prevalence and $\rho$ is a constant of proportionality.  This reflects perceptions regarding the risk of infection from traditional funerals which is assumed to be increasing in prevalence.  Sanitary burials incur a cost $C$ that reflects cultural pressure, so their payoff $u_S$ is simply
\begin{equation}
u_S = -C.
\end{equation}
We assume that this cost does not change significantly on the timescale of an Ebola outbreak as cultural norms often change at a generational pace in the absence of major social upheaval.  

The fraction of the population that prefers traditional burials at a given time is $f_T(t)$, and $f_S(t)$ is the fraction that prefer sanitary burials.  We assume that behavior change occurs due to a social imitation process where individuals compare the payoff from their own choice to the payoffs received by others, switching if the alternative appears to be sufficiently better.  This process can be represented by imitation dynamics \cite{hofbauer2003evolutionary}.  For traditional burials:  
\begin{equation}
\dot{f_T} = spf_Sf_T\cdot(C - \rho I)
\end{equation}
Individuals sample others at a rate $s$.  When an individual encounters another with a different strategy, they may adopt the other strategy with probability proportional to the difference between payoffs $p\cdot\Delta E = p\cdot(C- \rho I)$ where $p$ is the constant of proportionality for imitation (for $\dot{f_S}, p\cdot\Delta E = p\cdot(\rho I - C)$).
We allow individuals to change strategies randomly with a small probability $\mu$.  This approach is analogous to replicator-mutator dynamics in evolutionary biology.  The complete behavioral dynamics are
\begin{equation}
\dot{f_T} = s[pf_Sf_T\cdot(C - \rho I) + \mu\cdot(f_S - f_T)].
\end{equation}
The inner term can be interpreted as the possible outcomes when an individual considers whether or not to change their burial preference.  Either the individual encounters another who prefers sanitary burials and (possibly) switches their preference, or they may switch independently with a small probability.  Unfortunately, $s, p$,and $\rho$ cannot be estimated individually.  Instead we rescale the above using $\sigma = sp\rho$, $c = \frac{C}{\rho}$, and $m = \frac{\mu}{p\rho}$ giving
\begin{equation}
\label{eq:imitationdynamics}
\dot{f_T} = \sigma[f_Sf_T\cdot(c - I) + m\cdot(f_S - f_T)].
\end{equation}

\begin{figure}[!h]
    \begin{center}
    	\includegraphics[width=0.75\textwidth]{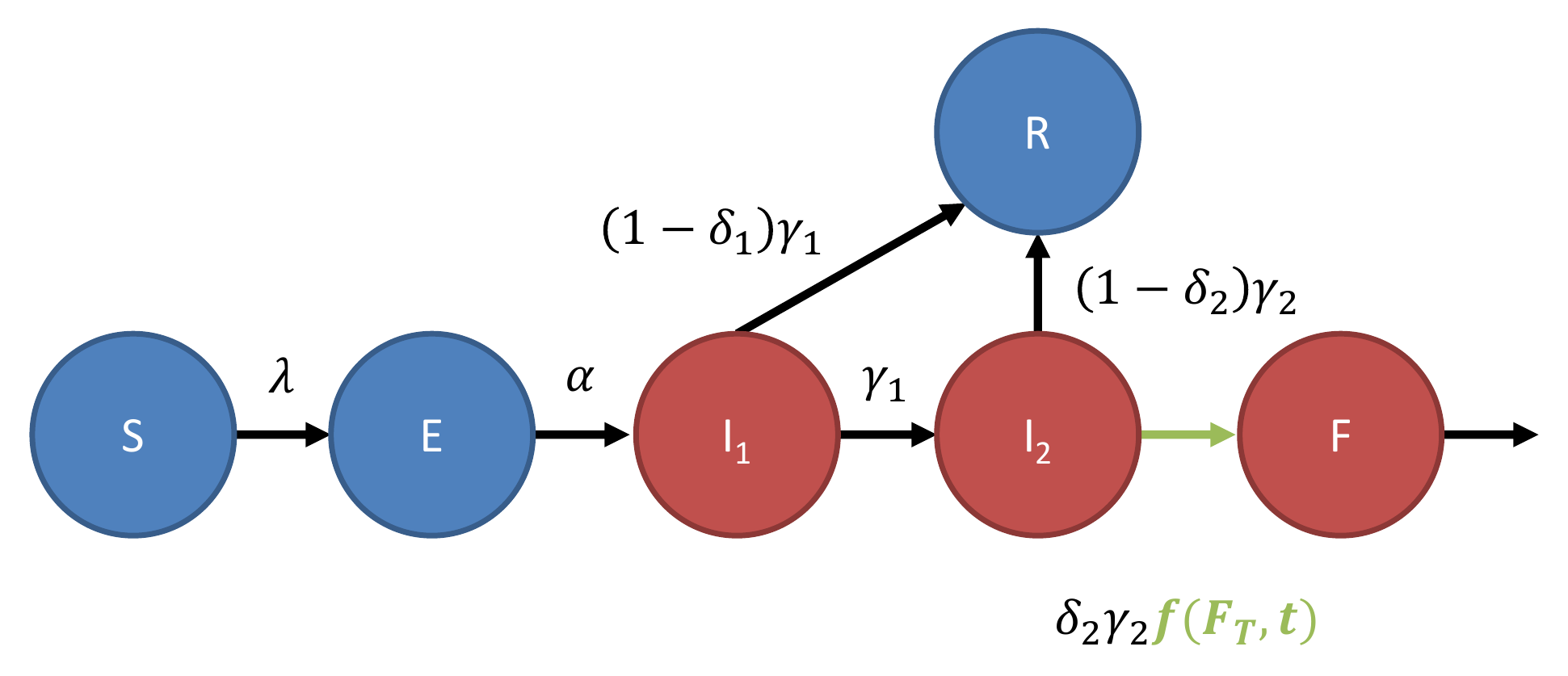}
    \end{center}
    \caption{{\bf Compartmental flow diagram of the Ebola transmission model}  Burial practices influence the transition between the second infected compartment $I_2$ and the funeral compartment $F$ (highlighted).}
    \label{fig:modeldiagram}
\end{figure}

\paragraph{Ebola virus transmission}
Clinically, EVD displays a multi-stage presentation with increasing severity and lethality over time \cite{ndambi1999epidemiologic,cdc2016ebola,world2016ebola}.  The initial stage includes non-specific febrile symptoms, while later stages progressively include diarrhea, vomiting, hemorrhage, and organ failure \cite{ndambi1999epidemiologic,cdc2016ebola,world2016ebola}.  Ebola virus remains viable in host fluids up to one week postmortem, enabling transmission by contact between uninfected individuals and infected cadavers \cite{prescott2015postmortem}.  We adapt the transmission model developed by Eisenberg et al. \cite{eisenberg2015ebola} by introducing the behavioral dynamics specified above.  Fig~\ref{fig:modeldiagram} depicts the compartmental structure of our model.  The full set of differential equations are 
\begin{align}
\begin{split}
\dot{S} &= -(\beta_1 I_1 + \beta_2 I_2 + \beta_F F)S \\
\dot{E} &= (\beta_1 I_1 + \beta_2 I_2 + \beta_F F)S - \alpha E \\
\dot{I_1} &= \alpha E - \gamma_1 I_1 \\
\dot{I_2} &= \delta_1 \gamma_1 I_1 - \gamma_2 I_2 \\
\dot{F} &= f_T \delta_2 \gamma_2 I_2 - \gamma_F F \\
\dot{R} &= (1-\delta_1)\gamma_1 I_1 + (1-\delta_2)\gamma_2 I_2 \\
\dot{f_T} &= \sigma[(1 - f_T)f_T\cdot(c - I) + m\cdot(1 - 2f_T)]
\end{split}
\end{align}

$S$ is the fraction of susceptible individuals, $E$ is the fraction exposed but not yet symptomatic, $I_1$ and $I_2$ are individuals in the first and second stage of infection, respectively, $F$ is the fraction who have died of EVD but have not yet been buried, and $R$ is the fraction who have recovered and are assumed to be immune.  Table \ref{tab:params} describes the parameters used in the above equations.  Note that $\beta_2$, $\beta_F$, and $\delta_1$ are derived parameters computed as follows: $\beta_2 = \beta_{21} \beta_1$, $\beta_F = \beta_{21} \beta_1$, and $\delta_1 = \delta/\delta_2$.  Like Eisenberg et al. \cite{eisenberg2015ebola} we estimate $\delta$ instead of $\delta_1$ or $\delta_2$ as the overall community mortality rate is more likely to be available based on the case fatality rate than stage-specific mortality.  We assume that all individuals practiced traditional burials before the outbreak, so $f_T(0) = 1$.  For our model fitting and simulations, we set $I_2(0) = 1/kN, S(0) = 1 - 1/kN$.  For Guinea, the index case was detected on December 26, 2013 \cite{who2015origins}.  For Liberia and Sierra Leone, we used the time of the first detected case.

\begin{table}[t]
    \centering
    \caption{{\bf Definitions of parameters used in the Ebola transmission model.}}
    \begin{tabular}{c | l | c | c }
        \bf Parameter & \bf Definition & \bf Units & \bf Source\\ \hline \hline
        $\beta_1$ & First stage transmission rate & person-days$^{-1}$ & Estimated \\
        $\beta_{21}$ & Second vs. first stage infectiveness ratio & unitless & Sampled \cite{yamin2015effect,francesconi2003ebola,dowell1999transmission,rivers2014modeling} \\
        $\alpha$ & Average incubation period$^{-1}$ & days$^{-1}$ & Sampled \cite{world2016ebola,cdc2016ebola} \\
        $\gamma_1$& First stage duration$^{-1}$& days$^{-1}$ & Sampled \cite{world2016ebola,ndambi1999epidemiologic}\\
        $\gamma_2$& Second stage duration$^{-1}$ & days$^{-1}$ & Sampled \cite{world2016ebola,ndambi1999epidemiologic}\\
        $\gamma_F$& Burial rate & days$^{-1}$ & Sampled \cite{world2016ebola,rivers2014modeling}\\
        $\delta$& Overall community mortality & unitless & Estimated \\
        $\delta_2$& Second stage mortality & unitless & Sampled \cite{ndambi1999epidemiologic} \\
        $k$& Population at risk & unitless & Estimated \\ 
        $N$ & Total population size & unitless & Fixed \cite{cia2016}\\
        $c$ & Private/social cost of sanitary burials & unitless& Estimated\\
        $\sigma$ & Imitation sampling/adoption rate & days$^{-1}$& Sampled\\
        $m$ & Random choice rate & days$^{-1}$ & Sampled \\ \hline
    \end{tabular}
    \label{tab:params}
\end{table}

The basic reproduction number for this model is
\begin{equation}
\Ro = \frac{\beta_1}{\gamma_1} + \frac{\beta_2 \delta_1}{\gamma_2} + \frac{f_T(0) \beta_F \delta_1 \delta_2}{\gamma_F}.
\end{equation}

This expression contains one term for each transmissible stage of EVD.  However, the burial transmission term is attenuated by the probability of a traditional burial at the disease free equilibrium.  As noted in \cite{hayashi2016effects}, the dependence of $\Ro$ on the behavioral initial condition can complicate its interpretation \cite{hayashi2016effects}.

\subsection*{Parameter estimation and sensitivity analysis}
We fit the model to cumulative incidence and mortality from the WHO situation reports (sitreps) \cite{who2016situation} using a hybrid approach.  While cumulative data can introduce estimation errors (primarily in variance estimates for deterministic models) \cite{king2015avoidable}, cumulative incidence and mortality were the only available data until relatively late in the outbreak.  Deriving incidence (and mortality) from the reported cumulative data would not be viable, as reporting errors resulted in apparent decreases in cumulative incidence and mortality.  Due to the large number of parameters in the model relative to the amount of available surveillance data, we fit a subset of parameters by numerical optimization, selecting others from plausible ranges derived from prior literature using Latin hypercube sampling (Appendix B).  

Given that only case and death data is available, it is possible that behavioral dynamics are not necessary to explain the epidemic trajectory.  In order to determine whether this is the case, we also fit a variant of the transmission model without behavior change and compare the residual error and Akaike Information Criterion (AIC) between models.

Our model and analyses are implemented in Python 2.7 using Numpy, Scipy, and Matplotlib.

\section*{Results and discussion}
\label{sec:results}
\subsection*{Parameter estimation}
\begin{figure}[!h]
    \centering
	    \includegraphics[width=\textwidth]{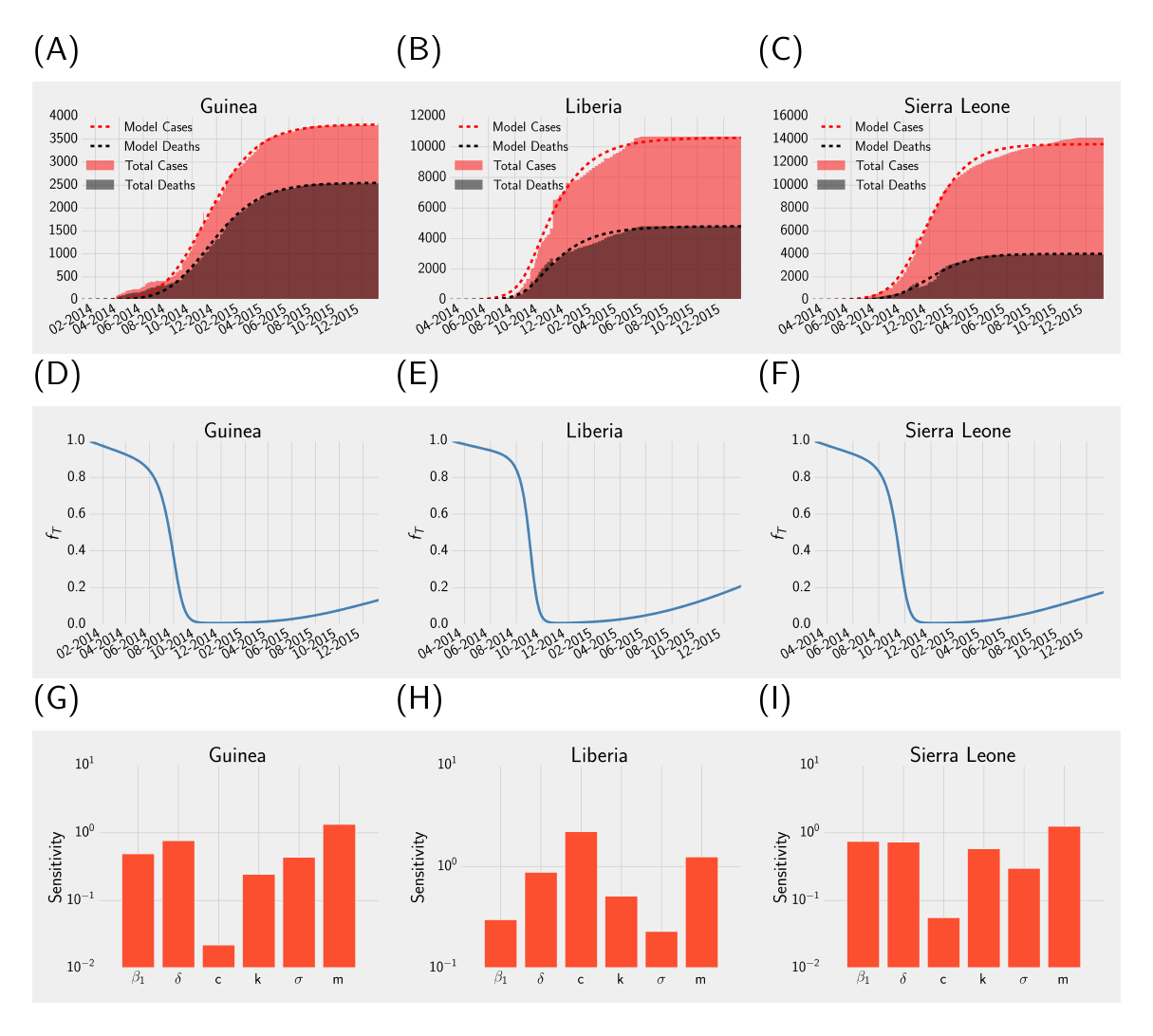}
    \caption{{\bf Burial dynamics model fit to cumulative cases and deaths in Guinea, Liberia, and Sierra Leone}  (A-C) compare the simulated cumulative cases and deaths (dashed lines) to the WHO sitrep data (shaded curves).  (D-F) show the predicted relative frequency of traditional burials from the best-fit models for each country.  (G-I) show the sensitivity coefficients for our directly estimated parameters ($\beta_1$, $\delta$, $c$, and $k$) as well as the remaining behavioral parameters ($\sigma$ and $m$).}  
    \label{fig:modelfits}
\end{figure}

Table \ref{tab:paramests} lists the best-fit values of $\beta_1$, $\delta_1$, $k$, and $c$  for each country as well as the LHS values for all other parameters corresponding to the best sample.  The fitted transmission and mortality parameters ($\beta_1$ and $\delta$) reflect the incidence and mortality trends in each country depicted in Fig~\ref{fig:modelfits}A--Fig~\ref{fig:modelfits}C.  In particular, Guinea has both the highest ratio of total deaths to total cases as well as the highest estimated mortality rate $\delta$, followed by Liberia and Sierra Leone.  The behavioral parameters $c$, $\sigma$, and $m$ were similar between the three countries.  This could reflect the fact that traditional burial practices are also similar within the region, so we would not expect the social/cultural pressure to hold a traditional burial to vary substantially.  In addition, the sampling rate $\sigma$ may reflect factors such as urbanization or interpersonal connectivity that influence how frequently any given individual would be exposed to information about burial practices.  The remaining best-fit sampled parameters are also relatively stable between countries with the exception of the burial rate in Liberia.  In general, this suggests that our selected transmission and behavioral parameters do inform differences between the outbreaks in each country.  Our estimates of $\Ro$ range from 1.4 to 1.56.  These values are relatively consistent between countries and are similar to other published estimates for Ebola \cite{althaus2014estimating,fisman2014early,eisenberg2015ebola}.  The best-fit trajectories for each country generally match the corresponding cumulative case and mortality data from the WHO sitreps (Fig~\ref{fig:modelfits}).  Our model is least accurate for Liberia due to the period of apparent linear growth in the sitrep data from December, 2014 to June, 2015.  It is not clear whether this phenomenon reflects actual transmission dynamics or is an artifact of more complete case detection catch-up in the later stages of the outbreak.  However, our model is still able to capture the majority of the outbreak growth dynamics as well as the final size.  

\begin{table}[h]
    \centering
    \caption{{\bf Best-fit parameter values for the full Ebola transmission model}}
    \begin{tabular}{c | c | c | c}
        \bf Parameter & \bf Guinea & \bf Liberia & \bf Sierra Leone \\ \hline \hline
        $\bm \beta_1$ & \bf{0.128} & \bf{0.106} & \bf{0.171} \\
        $\beta_{21}$ & 2.71 & 4.66 & 3.78 \\
        $\alpha$ & 0116 & 0.114 & 0.113 \\
        $\gamma_1$ & 0.158 & 0.363 & 0.189 \\
        $\gamma_2$ & 0.683 & 0.674 & 0.594 \\
        $\gamma_F$ & 0.952 & 0.319 & 0.651 \\
        $\bm \delta$ & \bf{0.667} & \bf{0.454} & \bf{0.296} \\
        $\delta_2$ & 0.825 & 0.820 & 0.887 \\
        $\bm k$ &  \bf{8.85$\times$10$^{-3}$} & \bf{8.68$\times$10$^{-3}$} & \bf{5.25$\times$10$^{-3}$} \\ 
        $\bm c$ & \bf{5.49$\times$10$^{-5}$} & \bf{3.92$\times$10$^{-4}$} & \bf{1.25$\times$10$^{-4}$} \\
        $\sigma$ & 9.77 & 9.54 & 8.93 \\
        $m$ & 10$^{-4.08}$ & 10$^{-4.16}$ & 10$^{-4.03}$ \\ \hline
        $\Ro$ & 1.46 & 1.40 & 1.56 \\ \hline
        Full model AIC & 19775.11 & 224872.35 & 206568.72 \\
        Reduced model AIC & 44535.18 & 607758.09 & 477497.34 \\ \hline
    \end{tabular}
    \begin{flushleft}
      Bold parameters were estimated by numerical optimization while all others were determined by LHS and correspond to the best sample. Model fit was computed using AIC (lower indicates better fit).  The reduced model AIC corresponds to a model without behavior change.
    \end{flushleft}
    \label{tab:paramests}
\end{table}

Sensitivity analysis indicates that the parameters we selected to fit directly do influence the quality of the model fit (Fig~\ref{fig:modelfits} G-I).  In addition, the behavioral parameters $\sigma$ and $m$ appear to have individual influences on the model output.  The scaled cost of traditional burials $c$ has the lowest sensitivity coefficient when estimated for Guinea and Sierra Leone.  Indeed, the range of plausible behavioral trajectories for the outbreak suggests that $c$ can be estimated or sampled within approximately a factor of ten (Appendix C).  Based on simulations using a range of $c$ close to the best fit, it appears that this parameter influences the general shape of the behavioral trajectory (Appendix C).  In particular, increasing $c$ increases the speed with which individuals revert to traditional burials following the outbreak as well as the steady state fraction of traditional burials. For practical purposes, it may be sufficient to establish which range yields dynamics that broadly match the observed outbreak.

Fig~\ref{fig:modelfits}D-Fig~\ref{fig:modelfits}F depict the predicted frequency of traditional burials over time from our best fit models.  For each country traditional burials decline rapidly between August and October, 2014.  Subsequently, our models predict that some individuals begin to revert to traditional burials once the outbreak has essentially ended.  The timing of the first behavioral shift is noteworthy.  While traditional burials decline somewhat during the growth phase of the outbreak, the largest change occurs after the peak simulated incidence.  In addition, this period corresponds to a phenomenon observed in Eisenberg et al. \cite{eisenberg2015ebola} regarding estimates of the reporting rate/population at risk parameter $k$: The best fit value varied depending on the amount of data used to fit the model.  In particular, $k$ was relatively stable using data up to September, 2014, but decreased by orders of magnitude once data from October and later was included.  Thus, it appears that transmission-related behavior change can explain some of the variation of the reporting rate/population at risk over time.  

\subsection*{Model comparison and forecasting}
\begin{figure}[!h]
    \centering
	    \includegraphics[width=\textwidth]{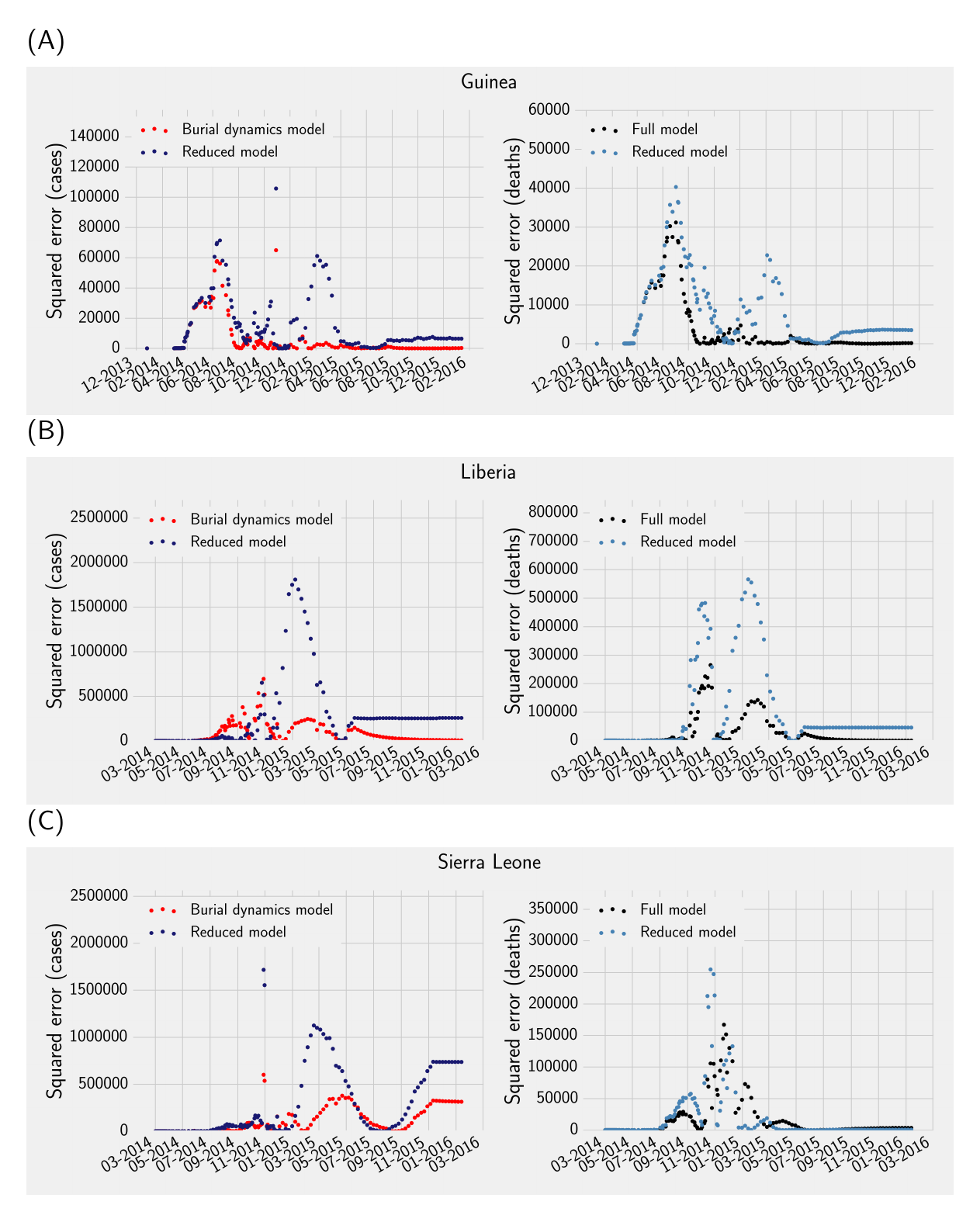}
    \caption{{\bf Residual error for the full and reduced models with respect to time}  The left and right panels show the error contribution of cumulative cases and cumulative deaths respectively.}
    \label{fig:fitcomparison}
\end{figure}

\begin{figure}[!h]
    \centering
	    \includegraphics[width=\textwidth]{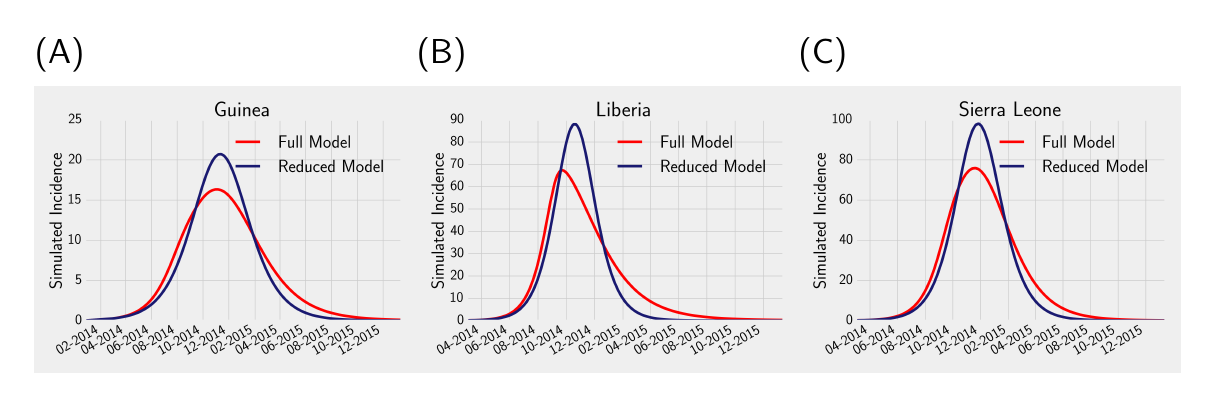}
    \caption{{\bf Simulated incidence from the best-fit full and reduced models}  (A) Guinea, (B) Liberia, (C) Sierra Leone.}  
    \label{fig:inccomparison}
\end{figure}

The reduced model without behavior change results in a worse fit both quantitatively and qualitatively.  Even after adjusting for an increase in the number of parameters, the full model AICs are substantially lower than those of the reduced model (Table \ref{tab:paramests}).  Fig~\ref{fig:fitcomparison} shows the squared residual error from both models for each time point in the sitrep data for each country.  Both models perform similarly early in the outbreak, but the reduced model cannot capture features of the data after incidence peaks.  This can also be seen by comparing the case and death trajectories from the reduced model to the sitrep data (Appendix C).  As a result, the best fit model without behavior change generally mis-predicts the final size of the outbreak.  This discrepancy is likely because the reduced model does not have a mechanism that can adjust transmission rates or the population at risk.  Fig~\ref{fig:inccomparison} compares the simulated incidence from the full and reduced models.  The full model generally produces a longer tailed incidence curve but lower peak incidence.  This is a consequence of the behavioral dynamics described above.  When traditional funerals decrease, the force of infection from funerals also decreases.  As a result, the susceptible population is depleted more slowly, allowing the outbreak to continue for a longer period of time.   

\begin{figure}[!h]
    \centering
	    \includegraphics[width=\textwidth]{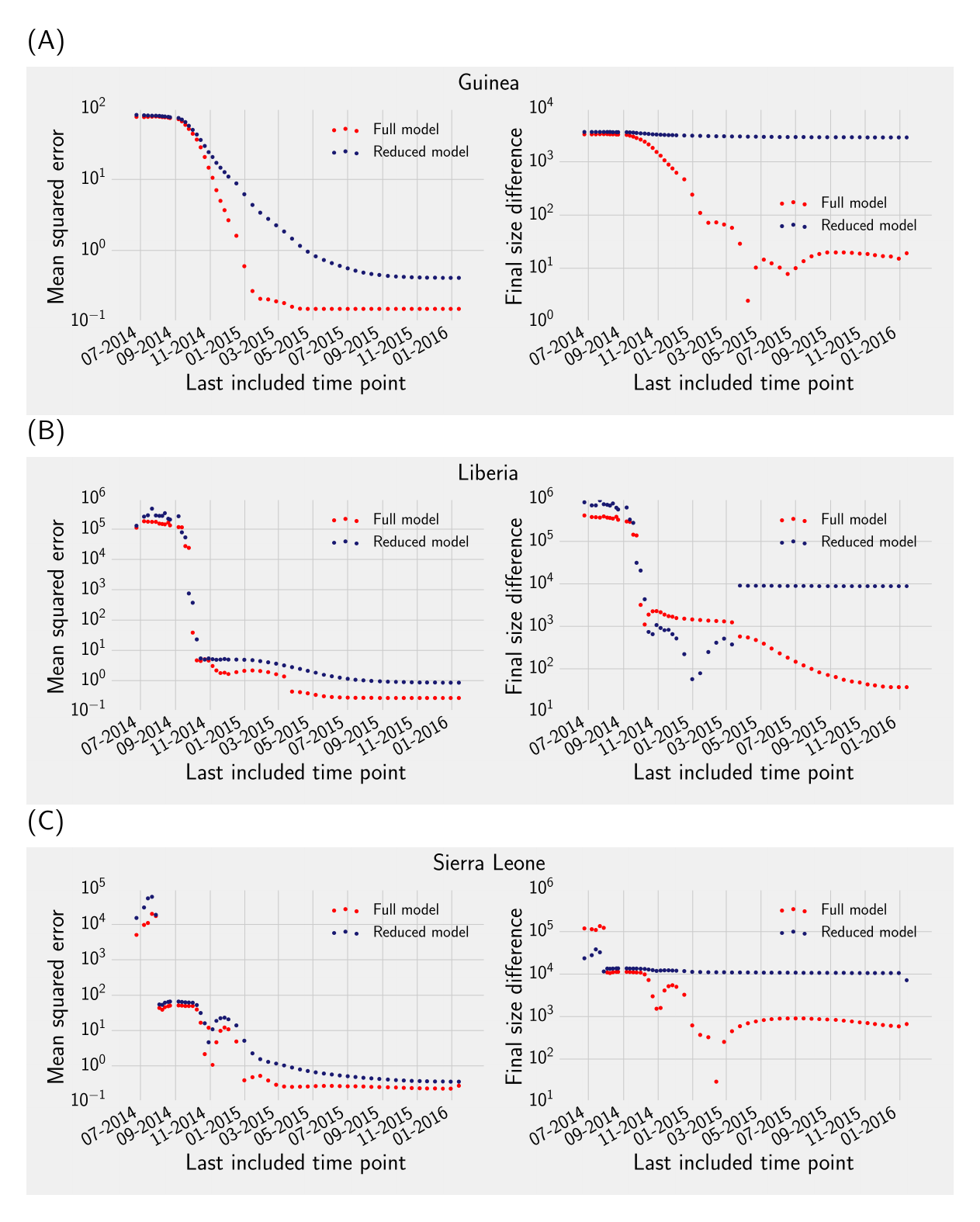}
    \caption{{\bf Comparison of forecasting accuracy between the full and reduced models}  Left panels (A, C, E) show the mean squared error for forecasts from the full and reduced model using data up to each included end point.  Right panels (B, D, F) compare the forecast final outbreak size from each model to the surveillance data final size.}
    \label{fig:forecastcomparison}
\end{figure}

We also evaluate the performance of our model by testing its forecasting accuracy relative to the reduced model.  To do this, we fit both models to truncated data from the outbreak then compute the mean squared error (MSE) for our output equations using the full outbreak data.  Thus we can test how effectively the model would have forecast the outbreak at different points in time.  Fig~\ref{fig:forecastcomparison} displays the full model's forecasting performance compared to the reduced model.  We are concerned with both how accurately the models predict the full time-course of the outbreak as well as whether they predict its final size.  While the forecasts are highly sensitive to the last data point included for fitting, the full model generally yields a lower MSE as well as a smaller difference between the predicted and actual final size (often by a full order of magnitude).  We do note however that neither model performs well until data from October or November, 2014 is included.  This corresponds to the inflection point in the outbreak data where incidence no longer increases exponentially.  The quality of projections from the full model do improve significantly after this point, further suggesting that the data contains a signal of behavior change.

\subsection*{Alternate scenarios}
\begin{figure}[!h]
    \centering
	    \includegraphics[width=\textwidth]{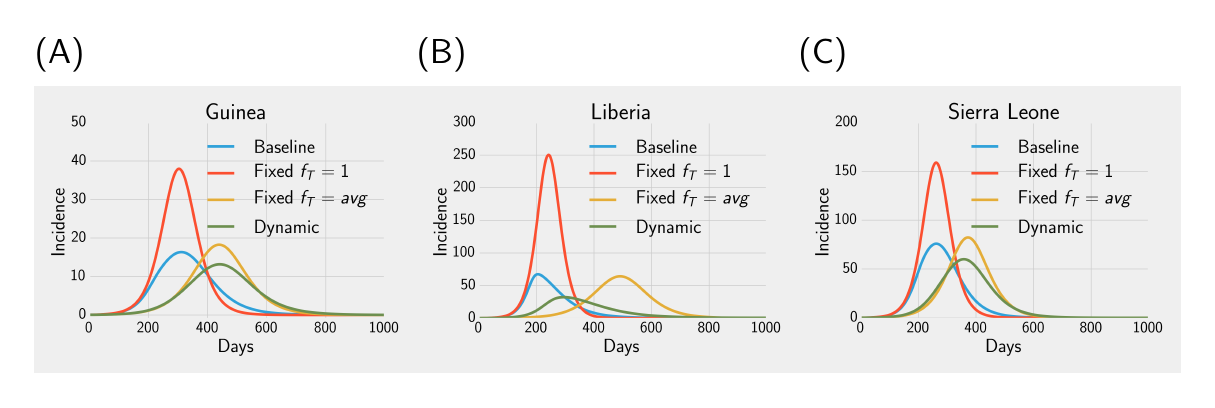}
    \caption{{\bf Simulated incidence in behavior change scenarios for each country}  Baseline curves represent incidence from the best-fit models shown in Fig~\ref{fig:modelfits}.  Total cases (by scenario) for Guinea (A) Baseline: 3825, $f_T=1$: 5836, $f_T=avg$: 4281, dynamic: 3647.  Total cases for Liberia (B) Baseline: 10628, $f_T=1$: 26920, $f_T=avg$: 15459, dynamic: 8461.  Total cases for Sierra Leone (C) Baseline: 13567, $f_T=1$: 19874, $f_T=avg$: 15148, dynamic: 12907}
    \label{fig:scenarios}
\end{figure}

We compare three hypothetical scenarios to evaluate the effect of adaptive burial practices as a control measure and to assess counterfactual outcomes for the outbreak.  In all three scenarios, we parameterize our model using the best fit values for each country and simulate outbreaks assuming a single initial infected individual.  The first scenario represents a worst case with respect to burial practices in which all funerals are traditional ($f_T = 1$) with no change over time.  The second scenario considers another fixed-behavior case -- the fraction of traditional burials is set to the average over the trajectory from the best-fit model.  This condition can be interpreted as an approximation of the actual behavioral dynamics assuming data collection does not capture the full trajectory.  For the final scenario, we set the initial fraction of traditional burials to its eventual steady state value from our fit to the 2014 outbreak and allow burial practices to change.  This represents a population that has previously experienced a large Ebola outbreak and adapted its burial practices accordingly, but practices continue to evolve.  Our model predicts that traditional burial practices do resume once the initial outbreak is over, so the new initial $f_T$ is above zero.  Fig~\ref{fig:scenarios} shows the incidence trajectories for each scenario as well as the best-fit (baseline) model.  We observe a similar phenomenon as in our model comparison.  Except for Liberia, both fixed-behavior scenarios have higher peak incidence than baseline, and generally symmetrical epidemic curves.  All other trajectories and final outbreak sizes were higher than the dynamic scenario, indicating that prior experience with an outbreak can reduce the overall magnitude of subsequent outbreaks.  Adaptation alone is not sufficient to prevent an outbreak, however, suggesting the need for additional intervention mechanisms.

\section*{Conclusion}

Behavior change during infectious disease outbreaks is frequently hypothesized or anecdotally reported, but rarely included explicitly in mathematical models.  We used evolutionary game theory to address the population dynamics of burial practices during the 2014 Ebola outbreak in West Africa.  In particular, our approach allowed us to assess the impact of adaptive behavior change on the scope of the outbreak.  Our full model reproduced the outbreak dynamics in Guinea, Liberia, and Sierra Leone, including an apparent decline in the force of infection and population at risk between August and October, 2014.  Behavior change, therefore, may have prevented the outbreak from attaining the magnitude of the more pessimistic early projections.  However, declining force of infection from burials may have also extended the duration of the outbreak.  These findings suggest that it will be important to consider the potential impacts of behavioral dynamics in future emerging disease scenarios.

A noteworthy feature of behavior-disease models is the potential to estimate behavioral parameters from surveillance data.  We found that the cost term for non-traditional burials could be directly estimated and the sampling and random choice rate could be determined by LHS.  While these parameters are abstractions of the true complex and stochastic determinants of burial practices and behavior change, they can provide insight into the relative degree of resistance to non-traditional practices.  Behavioral dynamics may explain observed changes in non-mechanistic parameters from fixed-behavior models.  Indeed, the behavioral trajectories from our best-fit simulations correspond to and may help to explain previously observed changes in the reporting rate/population at risk correction factor for the reduced model when it is fit to increasing amounts of data \cite{eisenberg2015ebola}.  Essentially, behavior change can alter the risk status of individuals, changing the size of the overall apparent population at risk.  The predicted decline in traditional burials provides evidence that behavior change may have reduced the force of infection, tipping the West Africa epidemic towards ending.  However, our model also projects an increase in traditional burials once the epidemic is over as the risk of infection due to funerals is once again low.  Thus while behavior change can contribute to the end of an outbreak, populations will not necessarily maintain their adherence with lower risk behaviors, suggesting a possible increase in risk for subsequent outbreaks in West Africa.

To account for the effect behavioral dynamics, infectious disease surveillance could be supplemented with time series of risk behaviors such as the fraction of traditional burials.  Behavior-disease models can use these sources with standard fitting methods.  Similarly models can be designed to include behaviors that are observed to change in response to outbreaks.  This may be particularly useful given that epidemics typically have similar early trajectories, characterized by a period of exponential growth.  During this period it is nearly impossible to discriminate between models of varying complexity with case and mortality data alone.  However, behavioral data streams may provide enough additional information to improve model selection or reduce uncertainty in estimates from a given model.  In the context of Ebola, collecting the relative frequency of burial types may have improved forecasting accuracy by signaling the reduction in transmission due to increasing sanitary burials.

Our forecasting results suggest that including behavior change may enable more accurate medium and long-term projections even using only  standard case and death surveillance data.  Our model gave reasonable predictions of the final size of the outbreak (within 8\% of the reported final size), although only once data from November, 2014 onward was included.  While final size information is often sought by policymakers early in an outbreak, these results underscore how surveillance data from this period may not be sufficient to provide an accurate estimate.  It is also difficult to determine ex ante whether a model's final size predictions are likely to be accurate.  However, a time varying force of infection due to behavioral adaptation is a plausible element in most outbreaks.  In particular, behavior change in our model leads to a reduced force of infection over time, and a lower final size as a result.  While still subject to substantial uncertainty during outbreak conditions, these estimates may be more plausible than higher estimates from models without behavior change.  

Our findings consistently indicate that burial practices changed significantly over the outbreak, however our model does not necessarily distinguish between changes in behavior due to interventions (e.g. increased burial team activity) and change due to social adaptation.  In particular the sanitary burial strategy represents an intervention as opposed to a burial type that existed prior to the outbreak.  Burial team deployment also increased as the scope of the epidemic became evident, so it is reasonable to assume that some of the predicted behavior change is capturing this phenomenon.  Still, burial teams' effectiveness depended on cooperation from local citizens.  As a result, our model can be interpreted as representing community-level adaptation to a behavioral intervention.  We also note while the qualitative features of our model's burial trajectories appear plausible, the exact degree of behavior change may be confounded by additional behavioral or intervention activity.  For example, Ebola treatment units (ETUs) were deployed in increasing numbers and individuals may have reduced their contact frequency while the epidemic was growing.  As current surveillance data is not likely to be sufficient to specify further behavioral mechanisms, future work may need to integrate alternative data sources such as anthropological studies.  Similarly, we omit healthcare transmission from our model to reduce its overall complexity.  This may bias our estimates of transmission terms somewhat as healthcare workers in Ebola treatment units may have experienced higher risk due to their frequent contact with late-stage patients.

In spite of these limitations, our model provides a platform to test hypothetical behavioral scenarios, which can seldom be studied experimentally.  For example, we compared the magnitude of simulated outbreaks between our best fit model and a combination of fixed and dynamic behavior conditions.  Both scenarios with dynamic burial behavior resulted in lower peak incidence than either fixed scenario.  This suggests that adaptation can provide protection even in populations that have not experienced prior outbreaks.  However, adaptation alone is not sufficient to prevent an outbreak from beginning.  In general, behavioral practices do not appear to change substantially until the outbreak is near or past its peak incidence.  This reflects the intuition that individuals may never react instantaneously to changing disease conditions.  Thus, prevention and rapid response to newly detected outbreaks are still crucial to successful control.  As emerging disease outbreaks increasingly occur in complex socio-political conditions, we argue that it is important to continue to develop methods that can provide mechanistic insights into behavioral processes as well as biological ones.

%
%
%

\end{document}